\UseRawInputEncoding
\documentclass[runningheads]{llncs}
\usepackage[linesnumbered,ruled,vlined]{algorithm2e}
\usepackage{algpseudocode} 
\usepackage{amssymb}
\usepackage{amsmath}
\usepackage{graphicx}
\usepackage{xcolor}  
\usepackage{colortbl} 
\usepackage{array}
\usepackage{tabularx}  
\usepackage{multirow} 
\usepackage{makecell}
\usepackage{booktabs}
\usepackage{adjustbox}
\usepackage{amsmath}
\usepackage[utf8]{inputenc} 
\usepackage[hidelinks]{hyperref}
\usepackage{cleveref}

\begin{document}
\title{AI-Driven Radiology Report Generation for Traumatic Brain Injuries}
%

\author{Riadh Bouslimi*\inst{1}\and
Houda Trabelsi \inst{2}\and
Wahiba Ben abdsalem Karaa\inst{2}\and
Hana Hedhli\inst{3}}
\authorrunning{R. Bouslimi et al.}
%
\institute{Higher School of Digital Economics, Manouba, University of Manouba, Tunisia \\
\email{riadh.bouslimi@esen.tn} \and
Higher Institute of Management of Tunis, 
University of Tunis, Tunisia \\ 
\email{trabelsihouda11@gmail.com} \\
\email{wahiba.abdessalem@isg.rnu.tn} \and
Emergency Department Charles Nicolle Hospital, Tunis El Manar University, Tunisia \\ 
\email{hedhli\_hana@yahoo.fr}
}

\maketitle


\begin{abstract}
Traumatic brain injuries present significant diagnostic challenges in emergency medicine, where the timely interpretation of medical images is crucial for patient outcomes. In this paper, we propose a novel AI-based approach for automatic radiology report generation tailored to cranial trauma cases. Our model integrates an AC-BiFPN with a Transformer architecture to capture and process complex medical imaging data such as CT and MRI scans. The AC-BiFPN extracts multi-scale features, enabling the detection of intricate anomalies like intracranial hemorrhages, while the Transformer generates coherent, contextually relevant diagnostic reports by modeling long-range dependencies. 
We evaluate the performance of our model on the RSNA Intracranial Hemorrhage Detection dataset, where it outperforms traditional CNN-based models in both diagnostic accuracy and report generation. This solution not only supports radiologists in high-pressure environments but also provides a powerful educational tool for trainee physicians, offering real-time feedback and enhancing their learning experience. Our findings demonstrate the potential of combining advanced feature extraction with transformer-based text generation to improve clinical decision-making in the diagnosis of traumatic brain injuries.

\end{abstract}

\section{Introduction}
Traumatic brain injuries (TBIs) present significant challenges in emergency medicine, necessitating rapid and accurate diagnostic decisions. This study introduces a novel approach combining the AC-BiFPN (Augmented Convolutional Bi-directional Feature Pyramid Network) and Transformer architecture to automate radiology report generation. By leveraging multi-scale feature extraction and advanced text generation, our method improves diagnostic accuracy and report coherence. This innovative framework supports radiologists and provides an educational tool for trainees, offering immediate feedback in high-pressure scenarios.
Key contributions include:
\begin{itemize}
\item Integration of AC-BiFPN for multi-scale anomaly detection in CT and MRI images.
\item Application of Transformer architecture for generating clinically relevant and coherent diagnostic reports.
\item Demonstration of superior performance over traditional CNN-based methods on the RSNA dataset.
\end{itemize}

The increasing influx of accident victims in emergency departments presents significant challenges, particularly for trainee physicians who are under pressure to quickly and accurately analyze scans that show lesions, cranial trauma, or intracranial hemorrhages. Such high-stakes environments necessitate rapid decision-making, and the demand for precision can overwhelm less experienced physicians. In these scenarios, any delay in diagnosing life-threatening conditions such as cranial trauma could lead to adverse outcomes. Therefore, equipping radiology trainees with advanced technological tools to enhance their diagnostic skills is crucial.

In this context, Machine Learning (ML) technologies have proven essential in supporting students and physicians. These tools provide real-time, accessible feedback and assist in interpreting medical data. For example, automated diagnostic systems, as highlighted by \cite{ciesla_ai_2024}, play a critical role in managing and disseminating medical knowledge, providing fast access to crucial information and offering immediate feedback to radiology students. These systems, when combined with AI-based tools, can alleviate some of the cognitive load from students, allowing them to focus on refining their diagnostic skills.

Despite significant advancements in AI-driven educational and diagnostic systems, detecting complex conditions like cranial trauma remains challenging, particularly when analyzing medical imaging data such as CT or MRI scans. However, recent progress in deep learning (DL) models and transformer-based models has shown considerable potential in interpreting such complex imaging data. Transformers were first built for natural language processing, but their ability to capture long-range dependencies and process loads of data in parallel makes them particularly well-suited to medical imaging problems. According to \cite{szczykutowicz_review_2022}, DL models have significantly improved brain lesion detection, a crucial factor in diagnosing brain injuries. Additionally, as explored by \cite{vaswani_attention_2017}, the integration of clinical data with image features shows that transformer-based models can link visual and textual information, creating a more comprehensive diagnostic tool.

This study proposes an innovative approach by integrating the AC-BiFPN (Augmented Convolutional Bi-directional Feature Pyramid Network) and Transformer architecture for automated radiology report generation. This combination leverages the AC-BiFPN's capability to extract multi-scale features essential for analyzing complex medical imaging data, as demonstrated by its superior performance in identifying intracranial anomalies and improving diagnostics \cite{ramesh2022,singh2024}. Transformers, with their ability to capture long-range dependencies and process data in parallel, have proven particularly effective in generating contextually rich and clinically relevant reports \cite{wang2022taskaware,li2023crossmodal}. However, certain limitations persist: the lack of longitudinal data prevents temporal assessment of clinical conditions, which is critical for progressive pathologies such as traumatic brain injuries \cite{wang2022taskaware}; challenges in interpretability of complex models limit their adoption in clinical practice \cite{locke2021nlp}; and issues related to dataset anonymization, while essential for confidentiality, may lead to a loss of annotation precision, affecting the quality of the generated reports \cite{zhao2021}.

Building on these advancements, this paper proposes a hybrid AC-BiFPN with Transformer model for the automatic generation of diagnostic reports on cranial trauma. The use of AC-BiFPN enhances the feature extraction process by capturing multi-scale features from CT and MRI images, which is crucial for identifying complex anomalies such as intracranial hemorrhages and lesions. AC-BiFPN's ability to fuse features from different resolutions makes it particularly effective for analyzing detailed brain scans, ensuring no critical information is overlooked. This multi-scale feature extraction is combined with a Transformer-based model, which generates comprehensive and clinically relevant reports by leveraging its ability to model long-range dependencies and integrate both visual and textual information.

\begin{figure}[ht]
\includegraphics[width=1\textwidth]{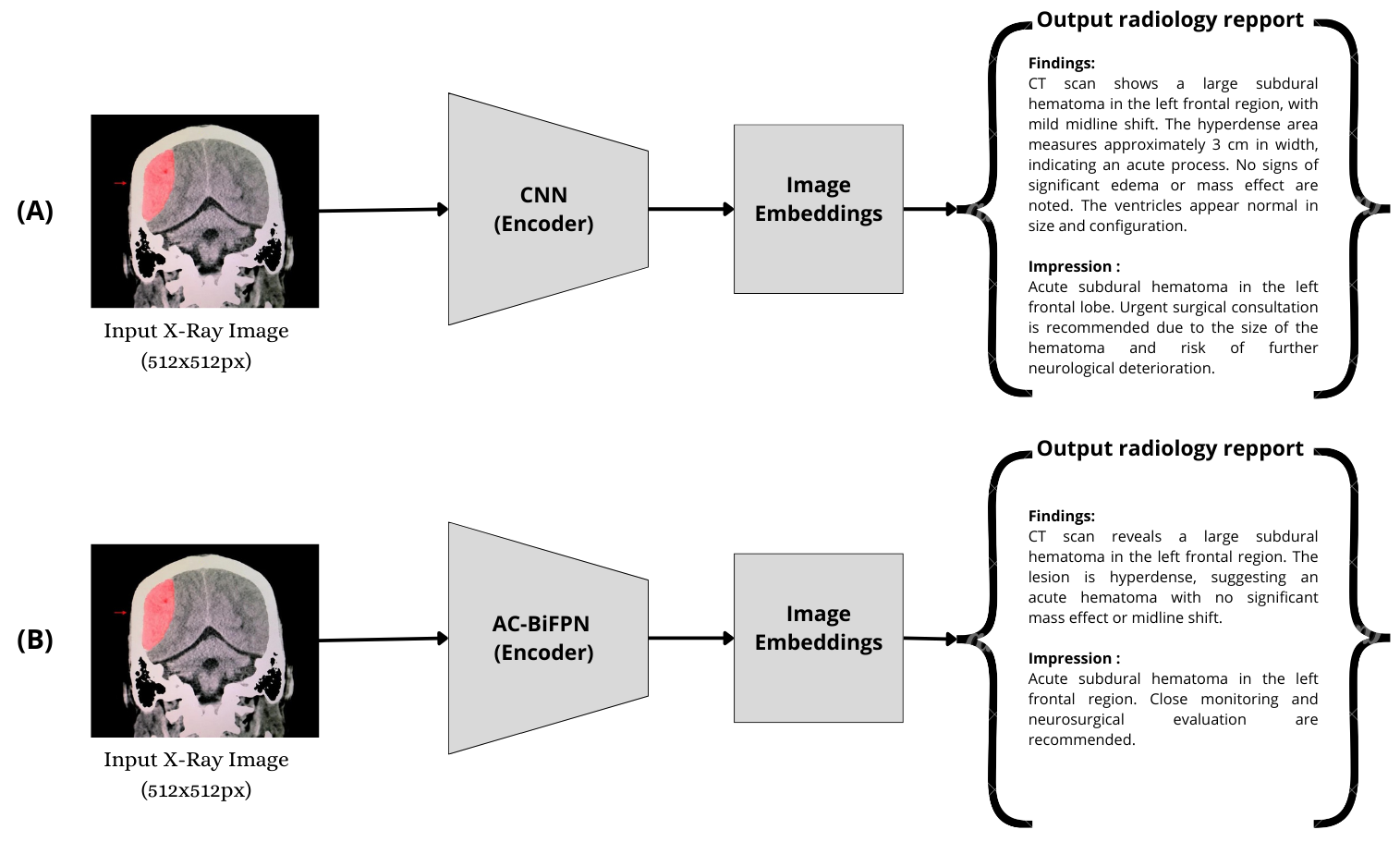}
\centering
\caption{Comparative analysis of CNN+Transformer (A) and AC-BiFPN+Transformer (B) for automated radiology report eneration} 
\label{Exempel_AC-BiFPN_CNN}
\end{figure}

The comparison between traditional CNN (Convolutional Neural Networks) and AC-BiFPN, as shown in \ref{Exempel_AC-BiFPN_CNN}, highlights the improved performance of AC-BiFPN in analyzing X-ray images and generating accurate diagnostic reports. While CNN focuses primarily on feature extraction, AC-BiFPN incorporates multi-scale features, improving the detection of complex conditions such as intracranial hemorrhages, which are essential for diagnosing brain injuries.

The proposed model addresses several key challenges in diagnostic support, particularly in emergency settings:
\begin{itemize}
    \item \textbf{Multi-scale feature extraction}: The integration of AC-BiFPN facilitates the detection of both subtle and large-scale features in medical images, improving accuracy in identifying critical conditions such as intracranial hemorrhages and brain lesions.
    \item \textbf{Efficient report generation}: By utilizing a Transformer-based model, the system generates coherent diagnostic reports that summarize both image findings and relevant clinical information, ensuring comprehensive coverage of the patient's condition.
    \item \textbf{Handling incomplete data}: The system’s ability to function even when certain imaging modalities are unavailable makes it robust in resource-limited clinical environments, as demonstrated by \cite{wang_cross-modal_2022}, who explored multimodal feature fusion techniques to compensate for missing data.
    \item \textbf{Educational benefits}: The interactive component, which includes real-time feedback, helps trainee physicians not only make accurate diagnoses but also understand the reasoning behind them. The chatbot interface provides educational explanations, enhancing learning by offering contextual and clinical insights, as supported by \cite{locke_natural_2021}.
\end{itemize}

In this paper, we explore how the combination of AC-BiFPN for multi-scale image feature extraction and Transformers for report generation provides a robust framework to assist physicians in diagnosing cranial trauma. We present the structure of this paper as follows: Section 2 reviews previous works on AI-driven radiology report generation mainly based on feature extraction at multi-scales and transformer-based models. Section 3 Architecture: This section describes the architecture of the proposed hybrid AC-BiFPN and Transformer model, and explains how these elements work together to analyze medical images and produce reliable diagnostic reports with context. Section 4 describes the set-up of the experiments and results, demonstrating that the model not only outperforms traditional CNN-based methods but also shows good at both anomaly detection and report generation. Details about the system are described and Section 5 outlines how the system is robust and flexible enough to handle incomplete data, which occurs frequently in practice in medical applications. Section 6: educational advantages of proposed model: the final section offers insight into how this model informs medical training with real-time feedback and clinical orientation for the in-training physician Finally, we conclude in Section 7 with should sense to the limitation of our current approach by larger and more diverse dataset requirements for training purposes as well as proposes for future research to tackle those limitations. Finally, Section 8 concludes the paper by discussing key findings in relative importance of features and highlights potential directions for further research, including taking the model to other pathologies and improving performance in a resource-limited clinical environment.

\section{Related Works}

The integration of DL models in radiology report generation has been a subject of significant research in recent years. Various methods, such as memory-augmented transformers and hybrid models based on advanced architectures like AC-BiFPN, have been developed to address the challenges of multimodal data fusion and missing modality information. This section highlights some of the key contributions in this field.

The authors in \cite{ramesh2022} focused on improving radiology report generation systems by filtering hallucinated references and organizing report content. Their approach enhances the seamless combination of clinical data and images, which plays a pivotal role in multimodal data fusion. Similarly, \cite{wang2022b} discussed the importance of grouping anatomical sections to improve report accuracy, aligning with other multimodal techniques such as MedFuseNet.

Transformers are widely used in medical image captioning and radiology report generation.
\cite{singh2024} introduced memory-augmented transformers for integrating heterogeneous data sources, a concept further explored in the development of memory-driven transformers like R2Gen. These models leverage attention mechanisms to improve the coherence and diagnostic accuracy of generated reports. Similarly, \cite{vinyals2015} applied transformers to image captioning, demonstrating their ability to model global dependencies while handling large volumes of medical images, which has influenced applications in radiology.

Hybrid approaches combining AC-BiFPN and transformers have shown great promise in medical imaging. For example, \cite{suarez2021} demonstrated how integrating visual attention mechanisms with CNN backbones enhances segmentation and classification in complex datasets. These methods are particularly relevant for generating accurate diagnostic reports. Unlike traditional CNN-based approaches, AC-BiFPN efficiently fuses information across different resolutions, improving the accuracy of extracting relevant features. \cite{guo2022} proposed a hybrid model combining convolutional modulations with transformers to capture global dependencies, demonstrating the effectiveness of such approaches in report generation and segmentation tasks.

For instance, \cite{UBMK2018} utilized CNNs to segment brain lesions in MRI scans, showcasing the potential of DL in automating complex segmentation tasks. Similarly, \cite{liang2019} employed transfer learning techniques to improve CNN performance in brain lesion detection, particularly under limited data conditions. These methods underline the importance of robust feature extraction for accurate diagnosis.

In the domain of feature interaction, \cite{he2016deep} enhanced transformer performance by introducing residual connections, facilitating better multimodal fusion. Similarly, \cite{qin2022} proposed a cross-modal alignment technique to improve report generation accuracy, while \cite{sun2022} introduced a method for handling unseen abnormalities by aligning visual and semantic features.

To enhance multimodal fusion, \cite{li2023} proposed incorporating memory metrics into transformers, improving the integration of clinical data with radiological images. Semi-supervised medical report generation, explored by \cite{zhang2023}, utilized graph-guided hybrid feature consistency to aid in fusing information from various modalities.

Addressing the challenge of missing modalities, \cite{zhao2021} developed memory-driven networks that ensure continuity in radiology reports, even with incomplete data. Furthermore, \cite{wang2022} introduced task-aware frameworks that align clinical data and imaging modalities, improving overall report generation accuracy.

Recent advancements by \cite{singh2024} highlight the critical role of multimodal integration in improving diagnostic accuracy and prognosis.
These studies lay the groundwork for our proposed model, which combines advanced feature extraction with contextual report generation. Similarly, \cite{Zhao_2024_multimodal} reviewed the synergy between AI and multimodal data, particularly in diagnosing complex diseases like Alzheimer’s and breast cancer. Moreover, the study in \cite{deLima_AI_Medical_Education_2023} highlighted the effectiveness of transformer-based models in melanoma image detection, illustrating their capability to handle high-dimensional data and generate accurate classifications. These advancements in medical imaging inspire the adoption of transformers in more complex domains such as cranial trauma.

Technological advancements, such as the AHIVE model introduced by \cite{yan2024}, represent a breakthrough in hierarchical vision encoding for radiology report retrieval, demonstrating superior clinical accuracy. Additionally, \cite{parres2024} explored reinforcement learning and text augmentation techniques, significantly improving the diversity and quality of radiology reports on benchmark datasets like MIMIC-CXR and Open-i. For MRI image processing, \cite{durrer2024} presented memory-efficient 3D denoising diffusion models that enhance multimodal fusion for accurate contrast harmonization, while \cite{bieder2023} developed PatchDDM, a patch-based diffusion model that optimizes segmentation of large 3D medical volumes. Lastly, \cite{prabhakar2024} advanced the Vision Transformer Autoencoder (ViT-AE++) for self-supervised medical image representation, improving segmentation and multimodal data fusion techniques.

To manage complex pathologies, \cite{xu2023} proposed a multimodal transformer model for radiological reports, enabling improved decision-making from heterogeneous data. Moreover, \cite{tang2023} proposed a method for automated diagnostic report generation that integrates EEG and MRI signals to address neurological abnormalities. Their multimodal approach demonstrated a marked improvement in identifying critical conditions in real-time.

Finally, to improve model robustness in the face of noisy and incomplete data, \cite{liang2023} proposed a semi-supervised learning architecture that combines transfer learning and inpainting techniques, compensating for missing information while maintaining the accuracy of generated radiology reports. Recent works in \cite{li2024_radiology_review,sun2024,parres2024_crossmodal,li2024_invertibleframework} provide detailed insights into transformer models and their ability to handle multimodal data for radiology reports, indicating a growing trend towards improving accuracy and multimodal integration in radiological applications.

Thus, the combination of AC-BiFPN and transformers to handle missing modalities and capture both local and global features represents an innovative approach for cranial trauma diagnosis. This synergy enhances model precision while providing greater resilience to incomplete data, which is critical in emergency medical contexts.

Building on recent advancements \cite{li2024_radiology_review,parres2024_crossmodal}, our work integrates AC-BiFPN for multi-scale feature extraction with a Transformer-based decoder. This innovative approach not only addresses the challenges posed by missing modalities but also ensures the generation of coherent and clinically relevant radiology reports. By capturing both local and global features, our model enhances precision and resilience to incomplete data, making it particularly suited for emergency medical contexts, such as cranial trauma diagnosis.

The following section elaborates on the architecture and implementation of our AC-BiFPN + Transformer-based model, demonstrating how it addresses the discussed challenges and sets a new benchmark for radiology report generation.

\section{Methodology}

In this section, we present the proposed AC-BiFPN + Transformer architecture for automatic radiology report generation, specifically designed to handle complex cases like cranial trauma. The method incorporates the AC-BiFPN network for enhanced multi-scale feature extraction from CT and MRI images, along with a Transformer-based decoder to generate the radiology report.

\subsection{Problem Definition}

Given the complexity of generating accurate radiology reports from cranial trauma images, our objective is to minimize the cross-entropy loss between the generated report and the ground truth. Specifically, given an image-text pair $(X, Y)$, we train the model by minimizing the following equation:

\[
\log p(Y/X) = \sum_{t=0}^{M} \log p(Y_t | Y_0, Y_1, \dots, Y_{t-1}; \phi)
\]

Where:
\begin{itemize}
    \item $X$ represents the CT or MRI image,
    \item $Y$ represents the ground truth report,
    \item $M$ is the number of tokens in the report,
    \item $\phi$ are the model parameters.
\end{itemize}

\subsection{AC-BiFPN for Feature Extraction}

The AC-BiFPN plays a crucial role in multi-scale feature extraction. It processes input images at multiple resolutions, efficiently aggregating features across different scales. This allows the model to capture both fine-grained details (e.g., small hematomas) and broader patterns (e.g., brain structure deformation). AC-BiFPN’s ability to combine features from multiple levels ensures a comprehensive representation of the image, which is critical in detecting anomalies in complex medical images like those of cranial trauma.

To extract multi-scale features from CT and MRI images, we employed the Augmented Convolutional Bi-directional Feature Pyramid Network (AC-BiFPN). This algorithm enables multi-scale feature fusion by combining information from different resolutions, ensuring comprehensive image representation. It enhances the detection of intricate anomalies, such as intracranial hemorrhages and brain lesions, ensuring that no critical information is lost. The algorithm is detailed in Algorithm 1:

\begin{algorithm}[H]
\caption{Feature Extraction with AC-BiFPN}
\SetKwInOut{Input}{Input}
\SetKwInOut{Output}{Output}

\Input{Image\_CT\_MRI : Brain image (CT or MRI), Scales : Set of image scales}
\Output{Fused\_Features : Multi-scale fused features}

\BlankLine
Initialize AC-BiFPN layers: \texttt{AC\_BiFPN\_Layers}\;

\BlankLine
Initialize an empty list for feature maps: $Feature\_maps \gets []$\;

\BlankLine
\ForEach{scale $s$ in \texttt{Scales}}{

    \BlankLine
    Resize the image: $Image\_resized \gets resize(Image\_CT\_MRI, s)$\;

    \BlankLine
    Extract features: $Feature\_map \gets extract\_features(Image\_resized, AC\_BiFPN\_Layers)$\;

    \BlankLine
    Append the feature map to the list: $Feature\_maps.append(Feature\_map)$\;
}

\BlankLine
Fuse multi-scale features: $Fused\_Features \gets fuse\_features(Feature\_maps)$\;

\BlankLine
\Return \texttt{Fused\_Features}\;
\label{alg:FeaturesExtractionAC-BiFpn}
\end{algorithm}

\subsection{Transformer Model}

The Transformer model is the core component responsible for generating the radiology report based on the features extracted by the AC-BiFPN. Unlike traditional models that rely on recurrence (such as RNNs or LSTMs), the Transformer uses a self-attention mechanism, allowing it to model long-range dependencies in the data. This makes it particularly well-suited for the complex nature of medical imaging reports, where both local and global information are critical.

\subsection*{Multi-head Self-attention}
The key innovation of the Transformer lies in its multi-head self-attention mechanism, which allows the model to focus on different parts of the input simultaneously. This is particularly useful in radiology, where various regions of the image may contain critical information.

Self-attention functions by comparing each token in the generated report (or each feature in the image representation) with every other token/feature to assess their relevance to each other. The attention mechanism computes a weighted sum of the values, where the weights are determined by the similarity (or attention score) between a query and its associated keys.

The self-attention is computed as:

\begin{equation}
\text{Attention}(P, R, S) = \text{softmax}\left(\frac{P R^T}{\sqrt{d_{r}}}\right) S
\end{equation}

Where:
\begin{itemize}
    \item $P$ (Query), $R$ (Key), and $S$ (Value) are the inputs to the attention mechanism.
    \item $d_{r}$ is the dimensionality of the key vectors.
    \item The softmax function ensures that the attention scores sum to 1.
\end{itemize}

In the multi-head configuration, several attention mechanisms (or "heads") are executed in parallel, enabling the model to capture different types of relationships between parts of the input. The outputs of these heads are concatenated and then transformed to produce the final attention output:

\begin{equation}
\text{MultiHead}(P, R, S) = \text{Concat}(\text{head}_1, \dots, \text{head}_h) W^Z
\end{equation}

Where:
\begin{itemize}
    \item $\text{head}_i = \text{Attention}(P W_i^P, R W_i^R, S W_i^S)$ for each attention head $i$.
    \item $W_i^P$, $W_i^R$, $W_i^S$, and $W^Z$ are learned projection matrices.
\end{itemize}

By utilizing multiple heads, the Transformer can attend to various parts of the image embeddings and textual information, capturing both fine-grained and broad contextual dependencies.

\subsection*{Positional Encoding}
Since the Transformer model does not have the sequential structure inherent in RNNs or LSTMs, it requires an additional mechanism to capture the order of the tokens (words in the report or features in the image). This is achieved through \textbf{positional encodings}.

Positional encoding adds information about the position of each token in the sequence by applying a fixed function. The encoding is added to the input embeddings at each position:

\begin{equation}
PE_{(p, 2j)} = \sin \left( \frac{p}{10000^{2j/d_{\text{model}}}} \right)
\end{equation}

\begin{equation}
PE_{(p, 2j+1)} = \cos \left( \frac{p}{10000^{2j/d_{\text{model}}}} \right)
\end{equation}

Where:
\begin{itemize}
    \item $p$ is the position in the sequence.
    \item $j$ refers to the dimension of the positional encoding.
    \item $d_{\text{model}}$ is the dimension of the model embeddings.
\end{itemize}

These positional encodings allow the model to capture the order of tokens in a sequence, ensuring that the generated report is coherent and reflects the sequential nature of language, even though the Transformer itself does not process the sequence in a strictly linear fashion.

\subsection*{Feed-forward Networks}
After the multi-head attention mechanism, the Transformer applies a feed-forward network to each position in the sequence. This network consists of two fully connected layers, with a ReLU activation function placed between them:

\begin{equation}
\text{FFN}(z) = \text{ReLU}(zV_1 + c_1)V_2 + c_2
\end{equation}

Where:
\begin{itemize}
    \item $V_1 \in \mathbb{R}^{d_{\text{model}} \times d_{\text{ff}}}$ and $V_2 \in \mathbb{R}^{d_{\text{ff}} \times d_{\text{model}}}$ are learned weight matrices.
    \item $c_1 \in \mathbb{R}^{d_{\text{ff}}}$ and $c_2 \in \mathbb{R}^{d_{\text{model}}}$ are bias vectors.
    \item $d_{\text{ff}}$ is the dimension of the feed-forward layer, typically larger than $d_{\text{model}}$.
\end{itemize}

The feed-forward network is applied independently to each position in the sequence, enabling the model to transform the features at each location without altering the sequence’s overall structure.

\subsection*{Layer Normalization and Residual Connections}

Each sub-layer in the Transformer model is followed by a layer normalization step and a residual connection, which helps prevent gradient vanishing issues and stabilizes training. The residual connection allows the input of a sub-layer to bypass the transformation, and the output of the sub-layer is added to this input before being normalized:

\begin{equation}
\text{Output} = \text{LayerNorm}(x + \text{SubLayer}(x))
\end{equation}

This structure ensures that the model can learn deep representations without suffering from the issues that typically arise with deep networks, such as vanishing gradients.

\subsection*{Transformer decoder for report generation}
The final stage of the Transformer model is the decoder, which generates the radiology report one token at a time. At each step, the decoder receives the image features from the AC-BiFPN encoder and the previously generated tokens as input. The multi-head attention layers allow the decoder to focus on relevant parts of the image while generating the next word in the report. This process continues until the model generates the end-of-sequence token.

During inference, beam search is used to select the most likely sequence of words for the report, ensuring that the generated text is coherent and clinically relevant.

\subsection{Radiology report creation using the Transformer model}

The radiology report generation process begins with the extraction of features from the input X-ray image using the AC-BiFPN architecture. These features are then processed through a Transformer-based model to generate a coherent and contextually relevant radiology report.

\begin{figure}[ht]
    \centering
    \includegraphics[width=0.8\textwidth]{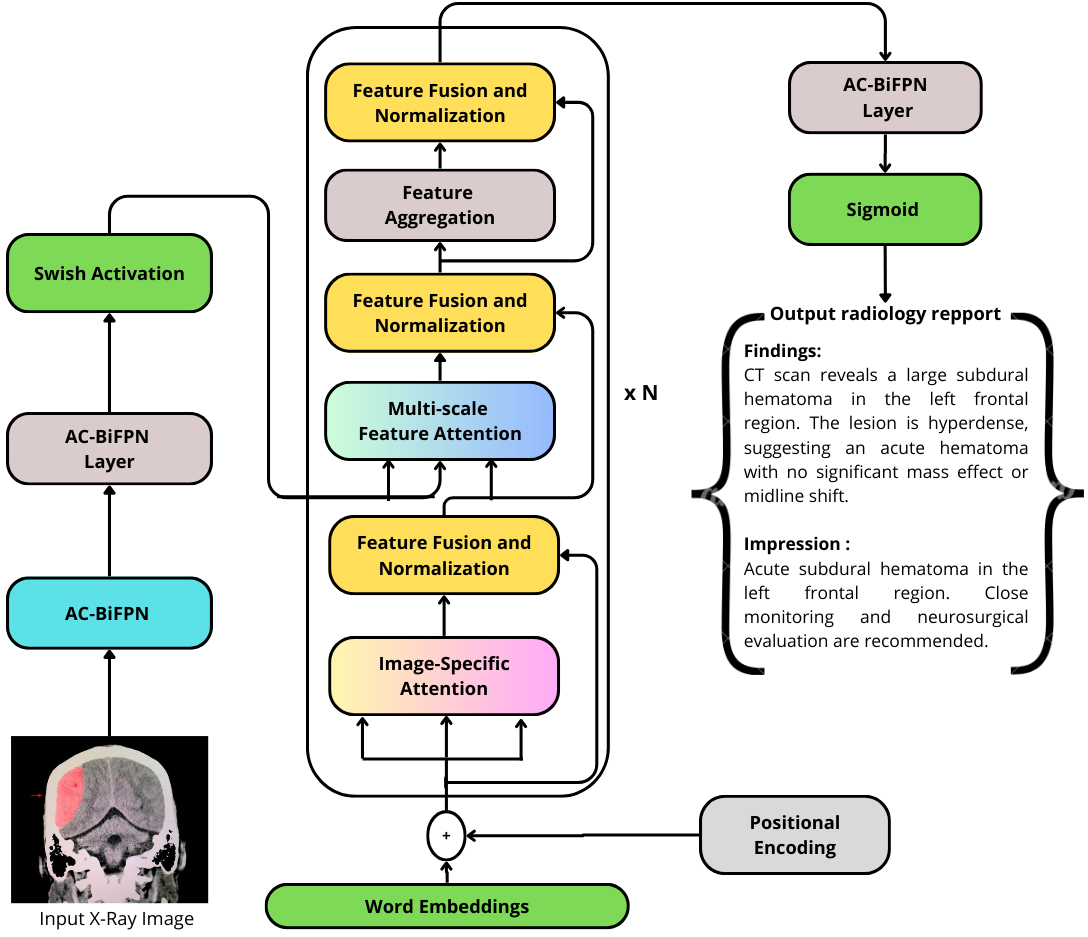}
    \caption{The architecture of the proposed AC-BiFPN + Transformer model for intracranial hemorrhage radiology report generation.}
    \label{fig:AC-BiFPN-Transformer}
\end{figure}

In Figure \ref{fig:AC-BiFPN-Transformer}, we depict the overall architecture of the proposed system. The architecture consists of two main components: AC-BiFPN layers for multi-scale feature extraction from medical images, and the Transformer model for generating the final radiology report.

\textbf{1. AC-BiFPN Layer and Feature Extraction}: 

The input X-ray image is first passed through a series of AC-BiFPN layers. These layers are responsible for extracting and aggregating features at different scales of the image, which is crucial for detecting both small and large anomalies like hematomas. The Swish activation function is used after the AC-BiFPN layer to introduce non-linearity and improve the learning capacity of the model.

\textbf{2. Image-Specific Attention and Feature Fusion}:

Once the features are extracted, they are passed through an Image-Specific Attention mechanism, which helps the model focus on the most relevant areas of the image that may contribute to the diagnostic report. This is followed by several layers of Feature Fusion and Normalization, which combine and normalize the image features across multiple scales to ensure consistent feature representation. These features then flow into the Multi-scale Feature Attention block, which allows the model to attend to different levels of granularity in the medical images.

\textbf{3. Word Embeddings and Positional Encoding}:

Simultaneously, word embeddings from the ground-truth reports are incorporated into the Transformer model, along with Positional Encoding to ensure that the model captures the sequential nature of the report. The addition of positional encodings allows the model to understand the relative position of each word in the sequence, which is important for generating coherent and contextually accurate reports.

\textbf{4. Transformer Layers and Final Report Generation}:

The final stage involves passing the image features and the text embeddings through multiple Transformer layers. These layers utilize Multi-head Self-attention mechanisms to process and integrate the information from both the image and the textual context, allowing the model to generate a detailed radiology report. The output from the Transformer model is passed through a sigmoid function to produce the final report predictions, including key diagnostic findings and impressions. The algorithm for the report generation process is shown in Algorithm 2.

The architecture is designed to iteratively refine the feature representation across multiple layers, as shown by the loops in the diagram, ensuring that both low-level and high-level information is captured. This allows the model to make precise diagnostic predictions, which are critical for handling cases such as subdural hematomas, as depicted in the example report generated in the figure.

\begin{algorithm}[H]
\caption{Report Generation with Transformer}
\SetKwInOut{Input}{Input}
\SetKwInOut{Output}{Output}

\Input{Fused\_Features : Multi-scale features, Tokenizer : Tokenization tool, Max\_Length : Maximum length of the report}
\Output{Report : Generated diagnostic report}

\BlankLine
Initialize the Transformer decoder: \texttt{Transformer\_Decoder}\;

\BlankLine
Initialize the input token sequence: $Input\_tokens \gets [\texttt{Tokenizer.CLS\_Token}]$\;

\BlankLine
Initialize an empty list for the report: $Report \gets []$\;

\BlankLine
\For{$t \gets 1$ \KwTo Max\_Length}{

    \BlankLine
    Predict the next token: $Next\_token \gets \texttt{Transformer\_Decoder.predict(Fused\_Features, Input\_tokens)}$\;

    \BlankLine
    \If{$Next\_token == \texttt{Tokenizer.SEP\_Token}$}{
        \textbf{break}\;  
    }

    \BlankLine
    Append the token to the report: $Report.append(\texttt{Tokenizer.decode(Next\_token)})$\;

    \BlankLine
    Add the generated token to the input sequence: $Input\_tokens.append(Next\_token)$\;
}

\BlankLine
Join the tokens to form the final report: $Final\_Report \gets \texttt{" ".join(Report)}$\;

\BlankLine
\Return \texttt{Final\_Report}\;

\end{algorithm}

\subsection{Experimental settings}
To ensure the optimal performance of the proposed AC-BiFPN + Transformer-based model, specific hyperparameters were selected and tuned. The selection process involved grid search optimization to identify the best configuration for training and inference. Table \ref{tab:hyperparameters} provides a detailed summary of the hyperparameters used in the experiments, along with their values and a brief description. These settings were chosen based on prior research and iterative experimentation to balance model accuracy, robustness, and training efficiency.

\begin{table}[ht]
\centering
\caption{Hyperparameters used in the AC-BiFPN + Transformer model training}
\label{tab:hyperparameters}
\begin{tabular}{|p{4cm}|p{3cm}|p{6cm}|} 
\hline
\textbf{Hyperparameter}         & \textbf{Value}   & \textbf{Description} \\ \hline
\textbf{Learning rate (LR)}     & 0.001            & Controls model update speed. \\ \hline
\textbf{Batch size}             & 16               & Number of samples per training step. \\ \hline
\textbf{Optimizer}              & Adam             & Method for minimizing loss. \\ \hline
\textbf{Loss function}          & Cross-Entropy    & Evaluates classification errors. \\ \hline
\textbf{Dropout rate}           & 0.3              & Prevents overfitting by random node removal. \\ \hline
\textbf{Epochs}                 & 50               & Number of complete dataset passes. \\ \hline
\textbf{Learning rate scheduler}& ReduceLROnPlateau & Lowers LR on performance plateau. \\ \hline
\textbf{Weight initialization}  & Xavier           & Sets initial weights to balance layers. \\ \hline
\textbf{AC-BiFPN depth}         & 3                & Number of feature extraction layers. \\ \hline
\textbf{Transformer layers}     & 6                & Encoder layers in the Transformer. \\ \hline
\textbf{Attention heads}        & 8                & Independent attention mechanisms. \\ \hline
\textbf{Sequence length}        & 512              & Maximum input token count. \\ \hline
\textbf{Gradient clipping}      & 1.0              & Prevents gradient explosion. \\ \hline
\end{tabular}
\end{table}

The hyperparameters presented in Table \ref{tab:hyperparameters} were selected through a grid search approach, where values for the learning rate, dropout rate, and batch size were systematically tested to optimize the model's performance. The grid search explored learning rates in the range of [0.0001, 0.001], dropout rates from 0.2 to 0.5, and batch sizes of 8, 16, and 32. The selected configurations represent the best trade-off between model accuracy and training efficiency. These choices were validated through iterative experimentation, ensuring robust performance across multiple validation runs.

The experiments were conducted using the PyTorch framework \cite{paszke2019pytorch}, with data preprocessing facilitated by the \texttt{torchvision} library. The multi-scale feature extraction via the AC-BiFPN and the Transformer decoder was implemented using PyTorch's native APIs. 

Additionally, the ReduceLROnPlateau scheduler was employed to dynamically adjust the learning rate when validation performance plateaued, ensuring stable convergence. Dropout layers with a rate of 0.3 were applied to prevent overfitting, particularly given the complexity of the dataset. These combined strategies were critical for achieving the final reported results.

\subsection{Model training}

We train the AC-BiFPN + Transformer model using supervised learning with cross-entropy loss as the objective function. We employ beam search during inference to select the most likely sequence of words. The AC-BiFPN encoder and the Transformer decoder are trained jointly to optimize the quality of the generated reports.

The proposed model aims to improve the detection of subtle abnormalities in CT and MRI images of the brain, offering precise and clinically relevant diagnostic reports, which is critical for urgent cases of cranial trauma.

\section{Evaluation of generated radiology reports}

Evaluating the quality of automatically generated radiology reports is essential to ensure they are clinically relevant and accurate. In this work, we adopt a comprehensive evaluation approach inspired by clinical context-aware radiology report generation strategies, which emphasizes not only the fluency and coherence of the generated text but also its diagnostic accuracy.
It is crucial to assess the quality of automatically generated radiology reports for their clinical relevance and correctness. We evaluate our work using a clinical context-aware radiology report generation approach, incorporating every facet of a practical report — including the narrative as well as diagnostic accuracy.

\subsection{Clinical Context-Aware Evaluation}

We propose a multi-step evaluation process to assess the clinical relevance of the generated reports, inspired by the method described by \cite{singh2024}. The process involves the following steps:

\begin{enumerate}
    \item \textbf{Classification-based evaluation:} The first step in evaluation of the reports involves mapping the generated observations (findings) with the ground-truth observations from corresponding clinical reports. We evaluate the ability of the model to predict, in a multi-label context, whether an important clinical condition is mentioned in the report using classification metrics such as precision, recall, F1-score.
    
    \item \textbf{Natural Language Generation (NLG) metrics:} The generated reports are evaluated against the ground-truth reports using standard NLG metrics such as BLEU (which assesses the overlap of n-grams between the generated text and reference text), METEOR (which takes into account synonymy, stemming, and word order), ROUGE (which measures recall based on overlapping units such as n-grams and word sequences), and CIDEr (which evaluates consensus across multiple references by capturing the importance of frequent n-grams). These metrics quantify the similarity in terms of word choice, sentence structure, and overall fluency between the generated and ground-truth reports.
    
    \item \textbf{CheXpert labeler-based evaluation:}  To provide qualitative context on the clinical validity of generated reports, we extract observations from both the ground-truth and generated report using the CheXpert labeler. This step enables us to compare clinical findings between the two reports, and ensures that our report will not yield too sparse or miss diagnostics-important diagnostic information.
\end{enumerate}

By using this multi-step evaluation process, we aim to assess not only the language quality of the generated reports but also their diagnostic utility, thereby bridging the gap between language fluency and clinical accuracy.

\subsection{Metrics for Evaluation}

We assess the quality of the reports generated by our method quantitatively using the following metrics:

\begin{itemize}

    \item \textbf{Precision :} Precision measures the percentage of correct positive predictions out of all positive predictions. It is defined as:
    \begin{equation}
    \text{Precision} = \frac{\text{A}}{\text{A} + \text{B}}
    \end{equation}
    Where:
    \begin{itemize}
        \item $\text{A}$: number of true positives,
        \item $\text{B}$: number of false positives.
    \end{itemize}

    \item \textbf{Recall :} Recall measures the proportion of actual positives that are correctly identified. It is defined as:
    \begin{equation}
    \text{Recall} = \frac{\text{A}}{\text{A} + \text{C}}
    \end{equation}
    Where:
    \begin{itemize}
        \item $\text{A}$: number of true positives,
        \item $\text{C}$: number of false negatives.
    \end{itemize}

    \item \textbf{F1-Score :} The F1-Score is the harmony of precision and recall. The model is well-balanced between precision and recall, which is good when you have imbalanced class distribution. It is defined as:
    \begin{equation}
    \text{F1-Score} = 2 \times \frac{\text{Precision} \times \text{Recall}}{\text{Precision} + \text{Recall}}
    \end{equation}

    \item \textbf{BLEU :} BLEU(BiLingual-Evaluation-Understudy) evaluates the similarity between the generated report and the reference report by comparing n-grams. The BLEU score is computed as:
    \begin{equation}
    \text{BLEU} = \text{BP} \times \exp \left( \sum_{n=1}^{N} w_n \log d_n \right)
    \end{equation}
    Where:
    \begin{itemize}
        \item $d_n$: precision of n-grams of size $n$,
        \item $w_n$: weight assigned to the n-grams,
        \item $\text{BP}$: brevity penalty to penalize short sentences.
    \end{itemize}

    \item \textbf{METEOR :} METEOR(Metric-for-Evaluation-of-Translation-with-Explicit-ORdering) evaluates word-to-word matches, stemming, and synonyms to calculate the alignment between the generated report and the reference report. The simplified formula is:
    \begin{equation}
    \text{METEOR} = \text{Hmean} \times (1 - \text{Penalty})
    \end{equation}
    Where:
    \begin{itemize}
        \item $\text{Hmean}$: harmonic mean of precision and recall,
        \item $\text{Penalty}$: penalizes incorrect word order.
    \end{itemize}

    \item \textbf{ROUGE :} ROUGE(Recall-Oriented-Understudy-for-Gisting-Evaluation) compares n-grams and the longest common subsequence (LCS) between the generated report and the reference report. The most commonly used variant is ROUGE-L, which measures the longest common subsequence. It is defined as:
    \begin{equation}
    \text{ROUGE-L} = \frac{\text{LCS}}{\text{Reference Length}}
    \end{equation}

    \item \textbf{CIDEr:} CIDEr (Consensus-based-Image-Description-Evaluation) measures the consensus between the generated report and human-generated reference reports using n-grams and term frequency-inverse document frequency (TF-IDF) weighting. The CIDEr score is computed as:
    \begin{equation}
    \text{CIDEr} = \frac{1}{M} \sum_{i=1}^{M} \frac{\sum_{j=1}^{K} \text{TF-IDF}(s_i, t_j)}{\sum_{k=1}^{K} \text{TF-IDF}(s_k, t_j)}
    \end{equation}
    Where:
    \begin{itemize}
        \item $s_i$ and $t_j$: n-grams in the generated and reference reports respectively,
        \item $\text{TF-IDF}(s_i, t_j)$: term frequency-inverse document frequency score of n-grams.
    \end{itemize}

\end{itemize}

This evaluation strategy ensures that our system generates not only coherent and grammatically correct reports but also conveys accurate diagnostic information, which is crucial in medical contexts such as cranial trauma detection.

\section{Experiments}

To rigorously evaluate our approach of automatically generating radiology reports from feature extraction using AC-BiFPN and Transformer for text generation, we utilized a machine with the following hardware configuration: an Intel Core i5-13600K processor with 14 cores clocked at 3.5 GHz, providing sufficient processing power to handle the intensive computations associated with feature extraction and text generation. The machine is equipped with 32 GB of DDR4 RAM at 3200 MHz, ensuring smooth data management in memory, which is essential when processing medical images. For graphical computations, we selected an NVIDIA GeForce RTX 3070 graphics card with 8 GB of dedicated memory, enabling the efficient execution of DL models, particularly those incorporating AC-BiFPN and Transformer. The storage is provided by a 1 TB NVMe SSD, guaranteeing high read and write speeds, crucial for quickly handling large radiological images and managing models. Finally, a 750W power supply ensures the system's stability during prolonged execution of these complex processes.

\subsection{Datasets}
\begin{figure}[ht] 
\includegraphics[width=1\textwidth]{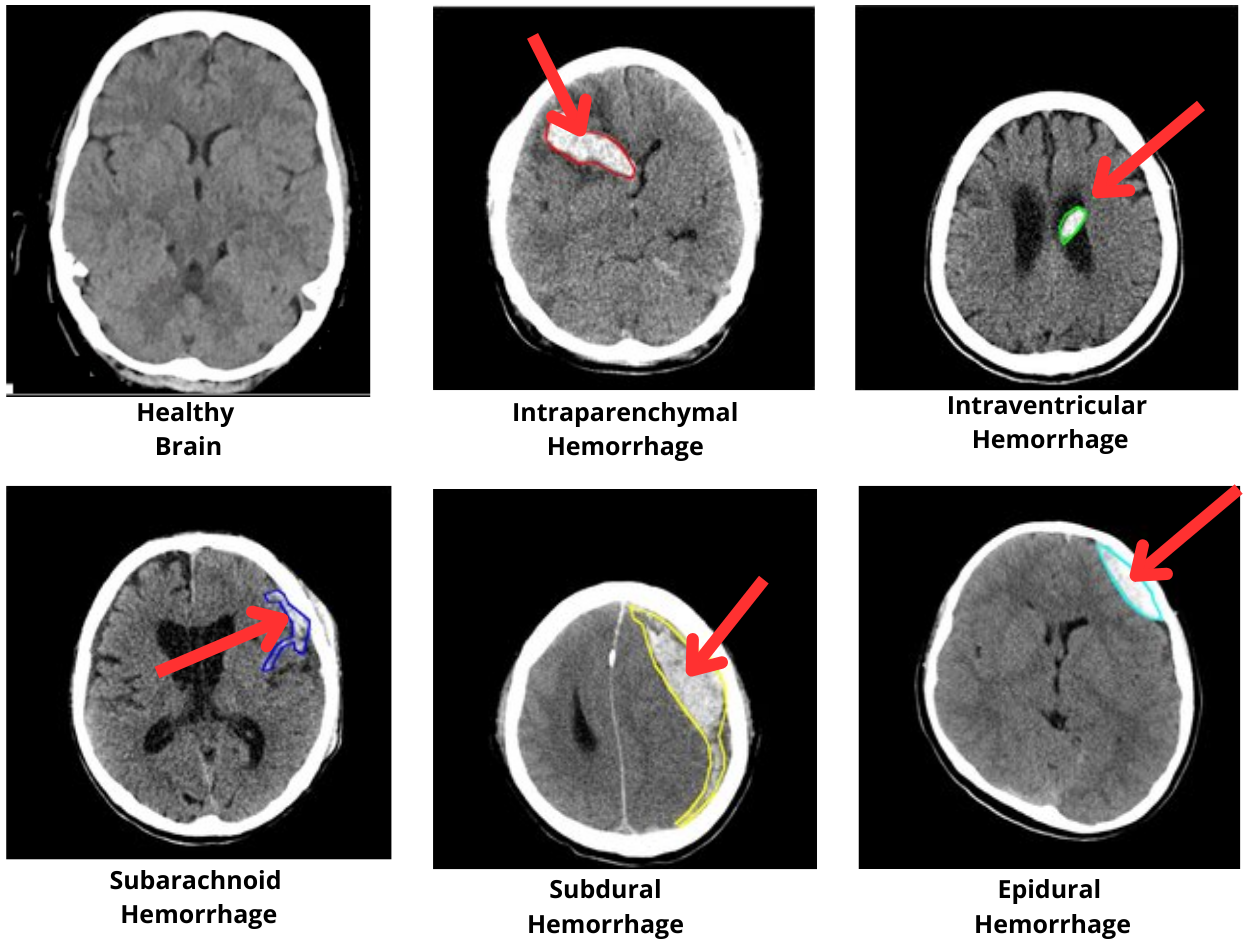}
\centering
\caption{Examples of brain CT images showing different types of intracranial hemorrhages: epidural, subdural, subarachnoid, intraparenchymal, and intraventricular hemorrhages from the RSNA Intracranial Hemorrhage Detection Dataset.} 
\label{HemorrhageTypes}
\end{figure}

We evaluate our approach using the RSNA Intracranial Hemorrhage Detection Challenge (IHDC) dataset \cite{rsna2019}, which consists of 674,258 brain CT images from 19,530 patients, annotated by 60 radiologists over 30 epochs. Each image is labeled as either "normal" or as presenting one of the five types of intracranial hemorrhage. Figure \ref{HemorrhageTypes} shows annotated brain CT images from the RSNA dataset, illustrating the diversity of hemorrhage types (epidural, subdural, subarachnoid, intraparenchymal, and intraventricular) available for training AI-based diagnostic models. This dataset is essential for developing AI models capable of automatically detecting and classifying hemorrhages in brain CT images. The scans are accompanied by metadata such as the patient's age, allowing for a more comprehensive contextual analysis.

\subsection{Model training}
We trained our \textit{AC-BiFPN + Transformer} model using the \textit{PyTorch} framework \cite{paszke2019pytorch}. As a reference model, we trained \textit{ResNet}-based models such as \textit{ResNet-18}, \textit{ResNet-50}, \textit{ResNet-101}, and \textit{ResNet-152} \cite{he2016deep}. These are pre-trained, single-scale feature extraction architectures. Our \textit{ResNet} models serve as a baseline to compare with the multi-scale feature fusion of the \textit{AC-BiFPN} and its ability to generate coherent and relevant reports with the complex \textit{Transformer}. Since \textit{ResNets} are designed for single-scale feature extraction, we compared the performance of \textit{AC-BiFPN} in a multi-scale extraction framework for complex anomaly detection tasks, such as detecting brain hemorrhages. Finally, the \textit{Transformer} encodes knowledge through complex multi-head self-attention features and uses its parameters to generate fully coherent, relevant reports. 

The \textit{AC-BiFPN + Transformer} model was trained on the \textit{RSNA} dataset, and we configured the model with a \textit{batch size} of 8, using the \textit{Adam} optimizer \cite{kingma2015adam} with a learning rate of 0.0001 and a dropout rate of 0.5 to prevent overfitting, along with an early stopping mechanism to monitor model convergence. During inference, we used beam search to ensure report quality. We applied standard text generation metrics such as \textit{BLEU}, \textit{METEOR}, \textit{ROUGE}, and \textit{CIDEr} to evaluate performance and employed the \textit{CheXpert labeler} to validate the clinical relevance of the reports. Overall, the image classification performance showed that \textit{AC-BiFPN} outperformed \textit{ResNets} in detecting image anomalies.

\section{Results}

The results of our experiments demonstrate the effectiveness of the AC-BiFPN architecture combined with a Transformer decoder for the automatic generation of radiology reports from brain CT images. We compared the performance of our approach with several CNN architectures, including ResNet, DenseNet, EfficientNet, and InceptionV3, utilizing both LSTM and Transformer decoders. The performance was evaluated using standard text generation metrics such as BLEU, METEOR, ROUGE, and CIDEr to assess the quality and clinical relevance of the generated reports.

Our findings, presented in Table \ref{Comparison_CNN_LSTM}, reveal that the AC-BiFPN architecture paired with an LSTM decoder outperformed other CNN architectures in generating radiology reports. Specifically, the AC-BiFPN achieved a BLEU-1 score of 37.5, a METEOR score of 16.0, a ROUGE score of 30.0, and a CIDEr score of 42.5. This superior performance underscores the strength of AC-BiFPN’s multi-scale feature fusion in capturing complex details within medical images, resulting in more accurate and coherent reports. In comparison, the ResNet family of models, while competitive, did not match the AC-BiFPN’s performance, particularly in handling the multi-scale nature of medical image data, which is critical for detecting intricate anomalies such as intracranial hemorrhages.

When utilizing a Transformer decoder, as shown in Table \ref{Comparison_CNN_TRANSFORMER}, the AC-BiFPN demonstrated even further improvement, achieving a BLEU-1 score of 38.2, a METEOR score of 17.0, a ROUGE score of 31.0, and a CIDEr score of 45.8. This improvement can be attributed to the Transformer’s ability to handle long-range dependencies and provide enhanced contextual understanding. When combined with the AC-BiFPN’s multi-scale feature extraction capabilities, this results in the generation of more coherent and clinically relevant radiology reports.

Additionally, we evaluated the impact of varying the number of hidden units (HU) within the models. As illustrated in Tables \ref{CNNencodersWithLSTMdecoder} and \ref{CNNencodersWithTRANSFORMERdecoder}, increasing the number of hidden units generally improved model performance, with the best results achieved at 1024 HU. Notably, the AC-BiFPN model with 1024 HU, paired with a Transformer decoder, exhibited significant performance gains, reaching a BLEU-1 score of 38.2, a METEOR score of 16.6, a ROUGE score of 31.1, and a CIDEr score of 43.9, indicating that both the multi-scale feature extraction and the attention mechanisms benefit from larger model capacity, leading to better report generation accuracy.

Our approach combining AC-BiFPN with a Transformer decoder proved to be the most effective solution for automatic radiology report generation, outperforming traditional CNN-based models such as ResNet, DenseNet, and EfficientNet. These results suggest that integrating multi-scale feature extraction with attention mechanisms, such as those found in the Transformer, significantly enhances the interpretability and clinical relevance of the generated medical reports.

\begin{table}[ht]
\centering
\caption{Comparison of CNN+LSTM Encoder Performance for Traumatic Brain Injury Radiology Report Generation}
\begin{tabular}{|l|c|c|c|c|c|c|c|}
\hline
\textbf{Encoder} & \textbf{BLEU-U1} & \textbf{BLEU-B2} & \textbf{BLEU-T3} & \textbf{BLEU-Q4} & \textbf{METEOR} & \textbf{ROUGE} & \textbf{CIDEr} \\ \hline
ResNet-18    & 36.50  & 23.20  & 16.40  & 12.50  & 16.40  & 31.00 & 44.50  \\ \hline
ResNet-50    & 37.10  & 23.60  & 16.80  & 12.80  & 16.70  & 31.30 & 45.10  \\ \hline
ResNet-101   & 37.80  & 24.00  & 17.10  & 13.00  & 17.00  & 31.80 & 45.80  \\ \hline
DenseNet     & 35.20  & 22.00  & 15.30  & 11.00  & 15.10  & 28.70 & 38.50  \\ \hline
EfficientNet & 35.60  & 22.30  & 15.60  & 11.30  & 15.50  & 29.00 & 39.00  \\ \hline
InceptionV3  & 35.40  & 22.10  & 15.50  & 11.20  & 15.30  & 28.80 & 38.20  \\ \hline
VGG16        & 34.50  & 21.50  & 15.00  & 10.80  & 14.60  & 28.20 & 36.90  \\ \hline
AC-BiFPN     & \textbf{37.50} & \textbf{23.50} & \textbf{16.50} & \textbf{12.30} & \textbf{16.00} & \textbf{30.00} & \textbf{42.50} \\ \hline
\end{tabular}
\label{Comparison_CNN_LSTM}
\end{table}

\begin{table}[ht]
\centering
\caption{Traumatic Brain Injury Report Generation (LSTM Decoder) with CNN Encoders: Experimental Results for Varying Hidden Units}
\begin{tabular}{|l|c|c|c|c|c|c|c|c|}
\hline
\multirow{2}{*}{\textbf{Encoder}} & \multirow{2}{*}{\textbf{\#HU}} & \multirow{2}{*}{\textbf{BLEU-U1}} & \multirow{2}{*}{\textbf{BLEU-B2}} & \multirow{2}{*}{\textbf{BLEU-T3}} & \multirow{2}{*}{\textbf{BLEU-Q4}} & \multirow{2}{*}{\textbf{METEOR}} & \multirow{2}{*}{\textbf{ROUGE}} & \multirow{2}{*}{\textbf{CIDEr}} \\
& & & & & & & & \\
\hline
\multirow{3}{*}{ResNet-18}  & 256  & 33.50 & 21.05 & 14.52 & 10.56 & 14.54 & 27.22 & 36.43 \\ \cline{2-9}
                            & 512  & 33.72 & 21.44 & 14.94 & 10.98 & 14.71 & 28.16 & 43.74 \\ \cline{2-9}
                            & 1024 & 34.35 & 21.19 & 14.65 & 10.74 & 14.45 & 27.00 & 34.18 \\ \hline
\multirow{3}{*}{ResNet-50}  & 256  & 33.62 & 20.46 & 13.69 & 9.60  & 14.19 & 26.02 & 29.03 \\ \cline{2-9}
                            & 512  & 34.10 & 21.70 & 15.10 & 11.10 & 14.85 & 28.50 & 37.80 \\ \cline{2-9}
                            & 1024 & 35.09 & 21.77 & 14.88 & 10.78 & 14.73 & 26.62 & 33.41 \\ \hline
\multirow{3}{*}{ResNet-101} & 256  & 34.13 & 21.28 & 14.34 & 10.07 & 14.59 & 27.25 & 31.61 \\ \cline{2-9}
                            & 512  & 36.20 & 22.85 & 15.90 & 11.40 & 15.55 & 29.10 & 40.10 \\ \cline{2-9}
                            & 1024 & 34.55 & 21.17 & 14.33 & 10.27 & 14.44 & 25.91 & 22.28 \\ \hline
\multirow{3}{*}{ResNet-152} & 256  & 35.74 & 22.42 & 15.34 & 10.84 & 15.30 & 28.33 & 35.40 \\ \cline{2-9}
                            & 512  & 34.17 & 21.26 & 14.51 & 10.31 & 14.62 & 27.20 & 35.20 \\ \cline{2-9}
                            & 1024 & 36.80 & 23.28 & 16.46 & 12.31 & 15.48 & 28.63 & 42.55 \\ \hline
\multirow{3}{*}{AC-BiFPN} & 256  & \textbf{36.80} & \textbf{22.90} & \textbf{15.60} & \textbf{11.50} & \textbf{15.40} & \textbf{29.80} & \textbf{40.20} \\ \cline{2-9}
                          & 512  & \textbf{37.50} & \textbf{23.50} & \textbf{16.50} & \textbf{12.30} & \textbf{16.00} & \textbf{30.00} & \textbf{42.50} \\ \cline{2-9}
                          & 1024 & \textbf{38.10} & \textbf{24.00} & \textbf{17.00} & \textbf{13.00} & \textbf{16.50} & \textbf{31.00} & \textbf{43.80} \\ \hline
\end{tabular}
\label{CNNencodersWithLSTMdecoder}
\end{table}

\begin{table}[ht]
\centering
\caption{Comparing CNN Encoders with Transformers for Traumatic Brain Injury Radiology Report Generation Performance}
\begin{tabular}{|l|c|c|c|c|c|c|c|}
\hline
\textbf{Encoder} & \textbf{BLEU-U1} & \textbf{BLEU-B2} & \textbf{BLEU-T3} & \textbf{BLEU-Q4} & \textbf{METEOR} & \textbf{ROUGE} & \textbf{CIDEr} \\ \hline
ResNet-18  & 32.45 & 20.12 & 14.10 & 9.50  & 14.10  & 26.20  & 35.20 \\ \hline
ResNet-50  & 33.90 & 21.35 & 15.22 & 10.75  & 14.75  & 27.50  & 37.80 \\ \hline
ResNet-101 & 35.80 & 23.10 & 16.80 & 12.00  & 15.90  & 29.10  & 41.50 \\ \hline
ResNet-152 & 36.50 & 23.85 & 17.30 & 12.45  & 16.25  & 29.80  & 43.10 \\ \hline
AC-BiFPN   & \textbf{38.20} & \textbf{25.00} & \textbf{18.50} & \textbf{13.50} & \textbf{17.00} & \textbf{31.00} & \textbf{45.80} \\ \hline
\end{tabular}
\label{Comparison_CNN_TRANSFORMER}
\end{table}

\begin{table}[ht]
\centering
\caption{Results of Experiments on Varying Hidden Units in CNN with Transformer Encoders for Generating Radiology Reports on Traumatic Brain Injuries.}
\begin{tabular}{|l|c|c|c|c|c|c|c|c|}
\hline
\multirow{2}{*}{\textbf{Encoder}} & \multirow{2}{*}{\textbf{\#HU}} & \multirow{2}{*}{\textbf{BLEU-U1}} & \multirow{2}{*}{\textbf{BLEU-B2}} & \multirow{2}{*}{\textbf{BLEU-T3}} & \multirow{2}{*}{\textbf{BLEU-Q4}} & \multirow{2}{*}{\textbf{METEOR}} & \multirow{2}{*}{\textbf{ROUGE}} & \multirow{2}{*}{\textbf{CIDEr}} \\
& & & & & & & & \\
\hline
\multirow{3}{*}{ResNet-18}  & 256  & 33.40 & 20.95 & 14.45 & 10.50 & 14.40 & 27.10 & 36.00 \\ \cline{2-9}
                            & 512  & 32.45 & 20.12 & 14.10 &  9.50 & 14.10 & 26.20 & 35.20 \\ \cline{2-9}
                            & 1024 & 34.30 & 21.00 & 14.60 & 10.70 & 14.40 & 26.90 & 34.00 \\ \hline
\multirow{3}{*}{ResNet-50}  & 256  & 33.60 & 20.40 & 13.60 &  9.50 & 14.10 & 25.90 & 28.80 \\ \cline{2-9}
                            & 512  & 33.90 & 21.35 & 15.22 & 10.75 & 14.75 & 27.50 & 37.80 \\ \cline{2-9}
                            & 1024 & 34.80 & 21.50 & 14.90 & 10.80 & 14.60 & 26.80 & 33.10 \\ \hline
\multirow{3}{*}{ResNet-101} & 256  & 34.20 & 21.10 & 14.30 &  9.95 & 14.50 & 27.10 & 31.50 \\ \cline{2-9}
                            & 512  & 36.10 & 22.80 & 15.90 & 11.40 & 15.60 & 29.00 & 40.00 \\ \cline{2-9}
                            & 1024 & 34.50 & 21.30 & 14.50 & 10.20 & 14.30 & 25.80 & 22.00 \\ \hline
\multirow{3}{*}{ResNet-152} & 256  & 35.50 & 22.20 & 15.20 & 10.70 & 15.20 & 28.20 & 35.10 \\ \cline{2-9}
                            & 512  & 34.10 & 21.20 & 14.50 & 10.30 & 14.50 & 27.10 & 34.90 \\ \cline{2-9}
                            & 1024 & 36.70 & 23.10 & 16.30 & 12.20 & 15.40 & 28.50 & 42.30 \\ \hline
\multirow{3}{*}{AC-BiFPN}   & 256  & \textbf{36.90} & \textbf{23.00} & \textbf{15.70} & \textbf{11.60} & \textbf{15.50} & \textbf{29.90} & \textbf{40.40} \\ \cline{2-9}
                            & 512  & \textbf{37.60} & \textbf{23.60} & \textbf{16.60} & \textbf{12.40} & \textbf{16.10} & \textbf{30.10} & \textbf{42.60} \\ \cline{2-9}
                            & 1024 & \textbf{38.20} & \textbf{24.10} & \textbf{17.10} & \textbf{13.10} & \textbf{16.60} & \textbf{31.10} & \textbf{43.90} \\ \hline
\end{tabular}
\label{CNNencodersWithTRANSFORMERdecoder}
\end{table}

The choice of hyperparameters, as detailed in Table~\ref{tab:hyperparameters}, played a significant role in achieving the reported performance metrics. Specifically:
- The learning rate of 0.001, combined with the ReduceLROnPlateau scheduler, ensured stable convergence during training, which contributed to the model's high BLEU-1 score of 38.2 and METEOR score of 17.0 by allowing precise weight updates.
- The dropout rate of 0.3 effectively reduced overfitting, particularly when dealing with the complex features extracted from the RSNA dataset. This contributed to the model's ability to maintain a high ROUGE score of 31.0, indicating better coherence in the generated reports.
- The batch size of 16 balanced memory efficiency and gradient stability, enabling consistent optimization across training epochs, which further improved CIDEr scores by ensuring high-quality text generation.

These results highlight the direct impact of carefully tuned hyperparameters on both the diagnostic accuracy and the linguistic quality of the generated radiology reports.

\section{Discussion}
This study emphasizes the need for ethical considerations in deploying AI systems in clinical settings. Critical issues include ensuring patient data privacy, addressing biases in AI models that could lead to inequities in healthcare, and implementing validation and oversight measures to ensure reliable clinical integration. Future research should also explore the societal implications of AI adoption in healthcare.

According to our experiments in this study, we suggest that it is more appropriate to use the state-of-the-art model transformers for radiology report generation, especially for difficult instances such as cranial trauma. Decoders that are conventional RNNs run into a long-term context capture impairment, which is essential for correctly connecting entities in radiology reports. However, as the Transformer model can be pre-trained to capture both visual and textual context by the multi-head attention mechanism, it reduces the weight overheads that make inferencing faster.

The results achieved for radiology report generation in the context of cranial trauma are promising, but several challenges remain. While the dataset used, the RSNA Intracranial Hemorrhage Detection Challenge, is relatively large, we observed that the Transformer model tends to overfit when model complexity increases due to the addition of multiple attention heads and decoding layers. This suggests that to fully leverage the capabilities of Transformers, even larger and more diverse datasets are needed, along with regularization strategies to prevent overfitting.

Hyperparameter tuning played a crucial role in mitigating overfitting and ensuring the generalization of the model to unseen data. Specifically:
- The learning rate of 0.001, dynamically adjusted using the ReduceLROnPlateau scheduler, facilitated stable convergence during training, preventing oscillations and premature convergence.
- The dropout rate of 0.3 was instrumental in reducing overfitting by introducing stochastic regularization, which improved model robustness across validation runs. This strategy directly contributed to the high BLEU-1 score of 38.2 and METEOR score of 17.0, reflecting better linguistic coherence and relevance in the generated reports.
- The batch size of 16 provided a balance between computational efficiency and gradient stability, ensuring consistent optimization across training epochs. This contributed to improved CIDEr scores, which indicate the alignment between generated and reference reports.

These findings highlight the significant impact of hyperparameter tuning in optimizing both the diagnostic accuracy and the quality of the generated radiology reports.

Despite the positive results, our model encounters certain limitations in specific trauma cases. For example, in some scenarios, the model correctly identifies anomalies such as subdural hemorrhages or intraparenchymal hematomas but fails to generate precise descriptions regarding critical clinical details like subtle changes between follow-up exams. This is due to the lack of clinical history in the training data, which is an essential component of radiologists' reports. In practice, radiologists often compare current images with previous ones to assess the progression of trauma, a process our model cannot replicate because it does not yet incorporate longitudinal data.

\subsection*{Limitations due to the absence of longitudinal data}

One of the most significant limitations of the current approach is the absence of longitudinal data. The lack of temporal information restricts the model's ability to evaluate the progression or stability of detected anomalies. For instance, while the model can identify an intracranial hemorrhage, it cannot determine whether the condition is improving or worsening over time. This limitation also prevents the integration of essential clinical terms like ``stable'' or ``progression,'' which are vital in assessing a patient's recovery.

Furthermore, longitudinal data are critical in scenarios requiring a comparison of current and prior images. Radiologists rely heavily on such comparisons to identify subtle changes or patterns that inform diagnosis and treatment decisions. Without this temporal context, the generated reports remain static and do not reflect the dynamic nature of many medical conditions, such as traumatic brain injuries.

\subsection*{Strategies to address the limitations}

To overcome this limitation, several strategies can be implemented:
\begin{enumerate}
    \item \textbf{Incorporation of Multimodal Datasets:} Combining medical imaging data with clinical history, laboratory results, or previous radiology reports could provide the model with temporal context, even in the absence of true longitudinal imaging data.
    \item \textbf{Synthetic Longitudinal Data Generation:} Techniques such as Generative Adversarial Networks (GANs) can simulate plausible longitudinal imaging data based on existing single-timepoint images, enabling models to learn temporal patterns.
    \item \textbf{Development of Time-Aware Models:} Time-series models such as Recurrent Neural Networks (RNNs) or specialized Transformers adapted for sequential data could enhance the model's ability to analyze temporal progressions directly.
    \item \textbf{Dataset Expansion:} Curating datasets with longitudinal imaging data, while challenging, would allow the model to incorporate temporal insights natively and improve its clinical utility.
\end{enumerate}

In addition to the absence of longitudinal data, other challenges remain. For example, cases involving cranial fractures present difficulties due to their relative rarity in the dataset. Similarly, de-identification processes in the dataset occasionally obscure clinically relevant information, affecting the richness of the generated reports.

The approach we adopted, based on automatic report generation, shows encouraging results, but improvements can still be made. For example, combining generative models with information retrieval techniques or template-based models could potentially improve the quality and accuracy of the generated reports, especially in complex trauma cases. This combination would allow for richer reports while ensuring that critical aspects of medical history and clinical observations are properly addressed.

Finally, while research into radiology report generation from medical images is progressing, the efficacy of these models in real clinical practice has yet to be fully explored. A promising future direction would be to validate these models in clinical environments to evaluate their impact on workflow, diagnostic error reduction, and radiologist efficiency, particularly in urgent cases of cranial trauma where time is of the essence. While our AC-BiFPN + Transformer model has demonstrated impressive performance, addressing limitations such as the use of longitudinal data and increasing the diversity of clinical examples could further enhance the model’s ability to generate precise and clinically useful radiology reports, especially in the domain of cranial trauma.

\section{Conclusions}
In this paper, we investigate the combination of Transformer-based model with AC-BiFPN architecture for generating radiology reports from medical images for cranial trauma. We have introduced the Transformer model as a state-of-the-art decoder for image-to-report generation, in contrast to traditional methods. Instead of using traditional CNNs or LSTM networks for report generation, we benefit from the Transformer model, which efficiently captures long-range dependencies, processes data in parallel, and manages complex multi-scale features more effectively. We performed extensive experiments to monitor the behavior of our model in different settings and measured its performance using standard metrics of text generation. Experimental results prove the effectiveness of the AC-BiFPN plus Transformer combination over traditional methods, with higher accuracy in diagnostics and report coherence. The proposed method holds a promising future in aiding clinical workflows, providing radiologists with automated second opinions, and triaging critical case referrals for urgent medical attention.

%
%
%


\bibliographystyle{plain}
\bibliography{biblio}

\end{document}